\documentclass[12pt]{iopart}
\usepackage{latexsym}
\usepackage{amscd}
\begin{document}

 \def\lambdabar{\protect\@lambdabar}
\def\@lambdabar{%
\relax
\bgroup
\def\@tempa{\hbox{\raise.73\ht0
\hbox to0pt{\kern.25\wd0\vrule width.5\wd0
height.1pt depth.1pt\hss}\box0}}%
\mathchoice{\setbox0\hbox{$\displaystyle\lambda$}\@tempa}%
{\setbox0\hbox{$\textstyle\lambda$}\@tempa}%
{\setbox0\hbox{$\scriptstyle\lambda$}\@tempa}%
{\setbox0\hbox{$\scriptscriptstyle\lambda$}\@tempa}%
\egroup
}

\def\bbox#1{%
\relax\ifmmode
\mathchoice
{{\hbox{\boldmath$\displaystyle#1$}}}%
{{\hbox{\boldmath$\textstyle#1$}}}%
{{\hbox{\boldmath$\scriptstyle#1$}}}%
{{\hbox{\boldmath$\scriptscriptstyle#1$}}}%
\else
\mbox{#1}%
\fi
}
\def\msf{\hbox{{\sf M}}}
\def\psf{\hbox{{\sf P}}}
\def\Nsf{\hbox{{\sf N}}}
\def\Tsf{\hbox{{\sf T}}}
\def\Asf{\hbox{{\sf A}}}
\def\Bsf{\hbox{{\sf B}}}

\def\msfsim{\bbox{{\sf M}}_{\scriptstyle\rm(sym)}}
\newcommand{\mcsim}{ {\sf M}_{ {\scriptstyle \rm {(sym)} } i_1\dots i_n}}
\newcommand{\mcs}{ {\sf M}_{ {\scriptstyle \rm {(sym)} } i_1i_2i_3}}

\newcommand{\beqan}{\begin{eqnarray*}}
\newcommand{\eeqan}{\end{eqnarray*}}
\newcommand{\beqa}{\begin{eqnarray}}
\newcommand{\eeqa}{\end{eqnarray}}

 \newcommand{\suml}{\sum\limits}
\newcommand{\intl}{\int\limits}
\newcommand{\rvec}{\bbox{r}}
\newcommand{\xivec}{\bbox{\xi}}
\newcommand{\Avec}{\bbox{A}}
\newcommand{\Rvec}{\bbox{R}}
\newcommand{\Evec}{\bbox{E}}
\newcommand{\Bvec}{\bbox{B}}
\newcommand{\Svec}{\bbox{S}}
\newcommand{\avec}{\bbox{a}}
\newcommand{\nablav}{\bbox{\nabla}}
\newcommand{\nuvec}{\bbox{\nu}}
\newcommand{\bvec}{\bbox{\beta}}
\newcommand{\vvec}{\bbox{v}}
\newcommand{\jvec}{\bbox{j}}
\newcommand{\nvec}{\bbox{n}}
\newcommand{\pvec}{\bbox{p}}
\newcommand{\mvec}{\bbox{m}}
\newcommand{\evec}{\bi{e}}
\newcommand{\eps}{\varepsilon}
\newcommand{\la}{\lambda}
\newcommand{\rad}{\mbox{\footnotesize rad}}
\newcommand{\scr}{\scriptstyle}
\newcommand{\latens}{\bbox{\Lambda}}
\newcommand{\pitens}{\bbox{\Pi}}
\newcommand{\cm}{{\cal M}}
\newcommand{\cp}{{\cal P}}
\newcommand{\intd}{\intl_{\mathcal D}}
\renewcommand{\d}{\partial}
\def\rmi{{\rm i}}
\def\rme{\hbox{\rm e}}
\def\rmd{\hbox{\rm d}}

\title{Expressing the electromagnetic interaction energy}
\author{C. Vrejoiu }
\address{Faculty of Physics, University of Bucharest, 76900, 
 Bucharest-Magurele, Romania E-mail : cvrejoiu@yahoo.com}
\begin{abstract}
The interaction energy of a $(\rho,\jvec)$ distribution of electric charges 
and currents with an electromagnetic external field is expressed by the 
Cartesian components of the multipole tensors of the given distribution. 
Special attention is paid to the reduction of these tensors to the 
symmetric traceless ones. Although one uses the Cartesian tensor components 
in the explicit calculations, the final results are given in a consistent 
tensorial form.
\end{abstract}
\section{Introduction}
As is well known, a charged system given by the densities 
$\rho(\rvec,t)$ and $\jvec(\rvec,t)$  of electric charges and currents,
 localized in a finite domain 
$\mathcal D$ may be described by an infinite system of electric and magnetic 
multipoles. As the electromagnetic field associated to this distribution in the exterior of 
$\mathcal D$ may be expressed as a multipole expansion, the interaction 
of this system with an external electromagnetic field may be also expanded 
in terms of the multipole moments. In [1,2] the reduction of multipole 
Cartesian tensors is studied in the cases of electrostatic and magnetostatic 
fields. In [3-5] there are some attempts to give consistent multipole expansions in Cartesian 
coordinates in the dynamic case. In the present paper we  use such an approach 
by expressing the interaction energy of a charged system with an 
arbitrary electromagnetic external field.
 Our principal goal is to give general formulae for the 
interaction energy in a consistent tensorial form. Particular cases and 
applications and, particularly, the contributions of the toroidal moments 
 are given in the related literature.

\section{Multipole expansion of the interaction energy}
Let  the interaction energy 
\begin{equation}\label{wint}
W_{\scriptstyle  \rm{int}}=\intd\left[\rho(\rvec,t)\Phi_{\scriptstyle \rm{ext}}
(\rvec,t)-\jvec(\rvec,t)\cdot\Avec_{\scriptstyle \rm{ext}}(\rvec,t)\right]
\rmd^3x.
\end{equation}
Let us the origin $O$ of the coordinates in the domain $\mathcal D$ and  the 
Taylor series of $\Phi_{\scriptstyle\rm{ext}}$ and $\Avec_{\scriptstyle \rm{ext}}$ 
introduced in the equation \eref{wint}:
\beqa\label{ws}
W_{\scriptstyle \rm{int}}&=&\suml^{\infty}_{n=0}\frac{1}{n!}\intd\rho(\rvec,t)
x_{i_1}\dots x_{i_n}\rmd^3x\left[\d_{i_1}\dots\d_{i_n}\Phi_{\scriptstyle
\rm{ext}}(\rvec,t)\right]_{\rvec=0}\nonumber\\
&-&\suml^{\infty}_{n=0}\frac{1}{n!}\intd j_i(\rvec,t)x_{i_1}\dots x_{i_n}
\rmd^3x\,\left[\d_{i_1}\dots\d_{i_n} A_{i{\scriptstyle \rm{(ext)}}}(\rvec,t)
\right]_{\rvec=0}.
\eeqa
Denoting in the following $(\Phi,\,\Avec)$ instead of 
$(\Phi_{\scriptstyle\rm{ext}},\,
\Avec_{\scriptstyle\rm{ext}})$, and introducing the $n$th order electric 
multipole tensor
\begin{equation}\label{pn}
\psf^{n)}(t)=\intd\rvec^n\rho(\rvec,t)\rmd^3x
\end{equation}
we may write
\beqa\label{wp}
W_{\scriptstyle\rm{int}}&=&\suml_{n\geq 0}\frac{1}{n!}\left[\psf^{(n)}(t)||
\nablav^n_0\right]\Phi\nonumber\\
&-&\suml_{n\geq 0}\frac{1}{n!}\intd x_{i_1}\dots x_{i_n}\,j_i(\rvec,t)
\rmd^3x\,\left[\d_{i_1}\dots\d_{i_n}A_i(\rvec,t)\right]_{\rvec=0}.
\eeqa
Here $\avec^n$ is the n-fold tensorial product $(\avec\otimes\dots\otimes
\avec)_{i_1\dots i_n}=a_{i_1}\dots a_{i_n}$ and denoting by
 $\bbox{\sf{T}}^{(n)}$ an $n$th order tensor,
$\bbox{\sf{A}}^{(n)}||\bbox{\sf{B}}^{(m)}$ is an $|n-m|$th 
order tensor with the components
\begin{displaymath}\left(\bbox{\sf{A}}^{(n)}||\bbox{\sf{B}}^{(m)}\right)_{i_1\dots i_{|n-m|}}=
\left\{\begin{array}{ll}A_{i_1\dots i_{n-m}j_1\dots j_m}B_{j_1\dots j_m}& \textrm{,  $n>m$}\\
A_{j_1\dots j_n}B_{j_1\dots j_n}&\textrm{,  $n=m$}\\
A_{j_1\dots j_n}B_{j_1\dots j_ni_1\dots i_{m-n}}&\textrm{,  
$n<m$}\end{array}\right. .\end{displaymath}
Let us 
\begin{equation}\label{wdef}
w^{(n)}=\intd x_{i_1}\dots x_{i_n}\,j_i(\rvec,t)\d^0_{i_1\dots i_n}A_i
\end{equation}
with
$$\d^0_{i_1\dots i_n}=\d^0_{i_1}\dots\d^0_{i_n},\;\;
\d^0_if(\rvec)=\d_if(\rvec)\big|_{\rvec=0}.$$
Considering the identity $\nablav\left[x_i\jvec(\rvec,t)\right]=j_i(\rvec,t)+
x_i\nablav\cdot\jvec(\rvec,t)$ and the continuity equation we have 
\begin{equation}\label{ceq}
j_i(\rvec,t)=\nablav\left[x_i\jvec(\rvec,t)\right]+x_i\frac{\d}{\d t}\rho
(\rvec,t).
\end{equation}
Using this equation in the equation \eref{wdef} and applying a procedure given in 
\cite{jack,castel}
we may write
\beqan
\fl w^{(n)}&=&-\intd x_i\jvec\cdot\nablav\left(x_{i_1}\dots x_{i_n}\right)\,\rmd^3
x\,\d^0_{i_1\dots i_n}A_i
+\intd x_{i_1}\dots x_{i_n}x_i\frac{\d\rho}{\d t}\,\rmd^3x\,\d^0_{i_1\dots i_n}A_i\\
\fl &=&-n\intd x_{i_1}\dots x_{i_{n-1}}x_ij_{i_n}\,\rmd^3x\,\d^0_{i_1\dots i_n}
\,A_i+\dot{\sf P}_{i_1\dots i_ni}\d^0_{i_1\dots i_n}\,A_i\\
\fl&=& -n\intd x_{i_1}\dots x_{i_{n-1}}\left(x_ij_{i_n}-x_{i_n}j_i\right)\,
\rmd^3x\,\d^0_{i_1\dots i_n}A_i-n\,w^{(n)}+
\left[\dot{\psf}^{(n+1)}||\nablav^n_0\right]\cdot\Avec
\eeqan
where some nul surface terms are considered  because $\jvec =0$ on $\d{\mathcal D}$ 
and  the super dot notation for the time derivatives is used. So, we get
\beqa\label{w}
\fl w^{(n)}&=&-\frac{n}{n+1}\intd x_{i_1}\dots x_{i_{n-1}}\left(x_ij_{i_n}
-x_{i_n}j_i\right)\rmd^3x\d^0_{i_1\dots i_n}A_i+\frac{1}{n+1}\left[\dot{\psf}
^{(n +1)}||\nablav^n_0\right]\cdot\Avec\nonumber\\
\fl&=&-\frac{n}{n+1}\eps_{ii_nk}\intd x_{i_1}\dots x_{i_{n-1}}\left(\rvec
\times\jvec\right)_k\rmd^3x\,\d^0_{i_1\dots i_n}A_i+\frac{1}{n+1}
\left[\dot{\psf}^{(n+1)}||\nablav^n_0\right]\cdot\Avec
\eeqa
By introducing the "vectorial product" $\Tsf^{(n)}\times \avec$ as the $n$th order tensor with 
the components
$$\left(\Tsf^{(n)}\times \avec\right)_{i_1\dots i_n}=\eps_{i_nij}{\sf T}_{i_1\dots 
i_{n-1}i}a_j,$$
and observing that, particularly,
$$\left(\bvec^n\times\avec\right)_{i_1\dots i_n}=\beta_{i_1}\dots
 \beta_{i_{n-1}}
\left(\bvec\times\avec\right)_{i_n},$$
we may use in the equation \eref{w} the definition of the $n$th order 
magnetic multipolar momentum \cite{castel}
\begin{equation}\label{mn}
\msf^{(n)}(t)=\frac{n}{n+1}\intd\rvec^n\times\jvec(\rvec,t)\,\rmd^3x
\end{equation}
such that the equation \eref{w} may be written as
\beqa\label{wfin}
\fl w^{(n)}&=&{\sf M}_{i_1\dots i_{n-1}k}\d^0_{i_1\dots i_{n-1}}\eps_{
ki_ni}\d_{i_n}A_i+\frac{1}{n+1}\left[\dot{\psf}^{(n+1)}||\nablav^n_0
\right]\cdot\Avec\nonumber\\
\fl &=& \left[\nablav^{n-1}_0||\msf^{(n)}\right]\cdot\left(\nablav_0\times\Avec\right)
+\frac{1}{n+1}\left[\dot{\psf}^{(n+1)}||\nablav^n_0\right]\cdot\Avec.
\eeqa
Using this last result we may write
\begin{equation}\label{wpol}
\fl W_{\scriptstyle\rm{int}}=\suml_{n\geq 0}\frac{1}{n!}\left[\psf^{(n)}||
\nablav^n_0\right]\Phi-\suml_{n\geq 1}\frac{1}{n!}\left[\dot{\psf}^{(n)}||
\nablav^{n-1}_0\right]\cdot\Avec
-\suml_{n\geq 1}\frac{1}{n!}\left[\nablav^{n-1}_0
||\msf^{(n)}\right]\cdot\Bvec.
\end{equation}
\section{The interaction energy and the reduced multipole tensors}
Using the notations from [1,3-5], the separation of the symmetric part 
$\msf^{(n)}_{\scriptstyle\rm{sym}}$ of the $n$th order magnetig multipole 
tensor $\msf^{(n)}$ is given by the formula
\begin{equation}\label{msim1}
{\sf M}_{i_1\dots i_n}={\sf M}_{\scriptstyle\rm{(sym)}i_1\dots i_n}
+\frac{1}{n}\suml^{n-1}_{\la=1}\eps_{i_{\la}i_nq}{\sf N}^{(\la)}_{
i_1\dots i_{n-1}q}
\end{equation}
where
\begin{equation}\label{msim2}
{\sf M}_{\scriptstyle\rm{(sym)}i_1\dots i_n}=\frac{1}{n}\left[
{\sf M}_{i_1\dots i_n}+{\sf M}_{i_ni_2\dots i_{n-1}}+\dots +
{\sf M}_{i_1\dots i_ni_{n-1}}\right]
\end{equation}
and
\begin{equation}\label{N}
N_{i_1\dots i_{n-1}}=\eps_{i_{n-1}ps}{\sf M}_{i_1\dots i_{n-2}ps}
\end{equation}
with the notation

$$f^{(\la)}_{i_1\dots i_n}=f_{i_1\dots i_{\la-1}i_{\la+1}\dots i_n}.$$
The symmetric tensor $\msf_{\scriptstyle\rm{sym}}$ is reduced to the 
symmetric traceless tensor $\bbox{\mathcal M}^{(n)}$ by the {\it detracer 
theorem} \cite{apple} which gives, with our notations,
\begin{equation}\label{detr}
\fl \msf_{\scr(\rm{sym})i_1\dots i_n}={\mathcal M}_{i_1\dots i_n}
-\suml^{[n/2]}_{m=1}\frac{(-1)^m(2n-1-2m)!!}{(2n-1)!!}\suml_{D(i)}
\delta_{i_1i_2}\dots\delta_{i_{2m-1}i_{2m}}\msf^{(n:m)}_{{\scr(\rm{sym})}
i_{2m+1}\dots i_n}
\end{equation}
where $[n/2]$ denotes the integer part of $n/2$, ${\sf M}^{(n:m)}_{{\scr (sym)}i_{2m+1}
\dots i_n}$ are  the components of the $(n-2m)$th-order tensor obtained from 
$\msfsim$ by the contractions of $m$ pairs of symbols $i$, and the sum over 
$D(i)$ is the sum over all the permutations of the symbols $i_1\dots i_n$ 
giving distinct terms. 
The symmetric traceless tensor $\bbox{\mathcal M}^{(n)}$ is given by 
\cite{cvsc}
\begin{equation}\label{mrond}
{\cal M}_{i_1\dots i_n}(t)=\frac{(-1)^n}{(n+1)(2n-1)!!}
\suml^n_{\la=1}\intl_{{\cal D}}r^{2n+1}\left[\jvec(\rvec,t)\times\nablav)\right]
_{i_{\la}}\d^{(\la)}_{i_1\dots i_n}\frac{1}{r}\rmd^3x.
\end{equation}

The equation \eref{detr} may be written as
\begin{equation}\label{mtr1}
{\sf M}_{\scriptstyle\rm{(sym)}i_1\dots i_n}={\mathcal M}_{i_1\dots i_n}
+\suml_{D(i)}\delta_{i_1i_2}\Lambda_{i_3\dots i_n}
\end{equation}
where $\bbox{\Lambda}^{(n-2)}$ is a symmetric tensor. Using the equation 
\eref{detr}, we may express 
$\bbox{\Lambda}^{(n-2)}$ by the formula
\beqa\label{lambda}
\fl \Lambda_{i_3\dots i_n}&=&\frac{1}{2n-1}{\sf M}_{(\scr sym)
qqi_3\dots i_n}\nonumber\\
\fl&+&\suml^{[n/2]}_{m=2}\frac{(-1)^{m-1}(2n-1-2m)!!}
{(2n-1)!!\,m}\suml_{D(i)}\delta_{i_3i_4}\dots \delta_{i_{2m-1}i_{2m}}
{\sf M}^{(n:m)}_{{\scr (sym)}i_{2m+1}\dots i_n}.
\eeqa
The reduction of the symmetric tensor $\psf^{(n)}$ is achieved by the relation 
\begin{equation}\label{ptr}
{\sf P}_{i_1\dots i_n}={\mathcal P}_{i_1\dots i_n}+\suml_{D(i)}
\delta_{i_1i_2}\Pi_{i_3\dots i_n}
\end{equation}
where the symmetric tensor $\bbox{\Pi}^{(n-2)}$ is defined in terms of the 
traces of the tensor $\psf^{(n)}$ by a relation similar to equation 
\eref{lambda}.
The symmetric traceless tensor $\bbox{\mathcal P}^{(n)}$ is given by 
the formula \cite{jansen}
\begin{equation}\label{Pi}
{\cal P}_{i_1\dots i_n}=\frac{(-1)^n}{(2n-1)!!}\intl_{{\cal D}}
\rho(\rvec,t)r^{2n+1}\nablav^n\frac{1}{r}\rmd^3x.
\end{equation}
Denoting by ${\mathcal W}_{\scriptstyle\rm{int}}$ the expression obtained from 
$W_{\scriptstyle\rm{int}}$ given by the equation \eref{wpol} by the
 substitutions 
$\psf^{(n)}\rightarrow \bbox{\mathcal P}^{(n)},\; \msf^{(n)}\rightarrow 
\bbox{\mathcal M}^{(n)}$ for all $n$, we  write
\beqa\label{wrond}
\fl  W_{\scriptstyle\rm{int}}&=&{\mathcal W}_{\scriptstyle\rm{int}}-
\suml_{n\geq 1}\frac{1}{n!n}\d^0_{i_1\dots i_{n-1}}\suml^{n-1}_{\la=1}
\eps_{i_{\la}kq}{\sf N}^{(\la)}_{i_1\dots i_{n-1}q}\left(\nablav_0\times
\Avec\right)_k\nonumber\\
\fl &-&\suml_{n\geq 1}\frac{1}{n!}\d^0_{i_1\dots i_{n-1}}
\left[\suml_{D(i)}\delta_{i_1i_2}\Lambda_{i_3\dots i_n}\right]\left(\nablav_0\times
\Avec\right)_{i_n}
-\suml_{n\geq 1}\frac{1}{n!}\left[\suml_{D(i)}\delta
_{i_1i_2}\dot{\Pi}_{i_3\dots i_n}\right]\d^0_{i_1\dots i_{n-1}}A_{i_n}\nonumber\\
\fl&+&\suml_{n\geq 0}\frac{1}{n!}\left[\suml_{D(i)}\delta_{i_1i_2}\Pi_
{i_3\dots i_n}\right]
\d^0_{i_1\dots i_n}\Phi.
\eeqa
After some straightforward calculations one obtains
\beqa\label{ww}
\fl W_{\scriptstyle\rm{int}}&=&{\mathcal W}_{\scriptstyle\rm{int}}+
\suml_{n\geq 2}\frac{n-1}{2n!}\left\{n\left[\Pi^{(n-2)}||
\nablav^{n-2}_0\right]\Delta_0\Phi\right.\nonumber\\
\fl &-&(n-2)
\left[\dot{\Pi}^{(n-2)}||\nablav^{n-3}_0\right]\cdot\Delta_0\Avec
-2\left[\dot{\Pi}^{(n-2)}||\nablav^{n-2}_0\right]\nablav_0\Avec
\nonumber\\
\fl&-&\left.\frac{2}{n}\left[\nablav^{n-2}_0|| \Nsf^{(n-1)}\right]\cdot
\left(\nablav_0\times\Bvec\right)
-(n-2)\left[\bbox{\Lambda}^{(n-2)}||\nablav
^{n-3}_0\right]\cdot\Delta_0\Bvec\right\}.
\eeqa
\par Using the expressions of the type \eref{lambda} for $\Lambda^{(n)}$ 
and $\Pi^{(n)}$  or using directly in the equation \eref{wpol} the {\it detracer theorem} from 
 \cite{apple} i.e. the relationships of the form \eref{detr}, 
 one obtains a detailed expression of $W_{\scr \rm int}$ 
in terms of the multipole tensors:
\beqa\label{wmp}
\fl W_{\scr \rm int}&=&{\mathcal W}_{\scr\rm int}+
\suml_{n\geq 2}\frac{1}{n}\left\{\suml^{[n/2]}_{m=1}\frac{(-1)^m(2n-1-2m)!!}{(2n-1)!!
2^mm!(n-2m)!}\left[-n\left(\psf^{(n:m)}||\nablav^{n-2m}_0\right)
\Delta^m_0\Phi\right.\right.\\
\fl &+&(n-2m)\left(\dot{\psf}^{(n:m)}||\nablav^{n-2m-1}\right)\cdot\Delta
^m_0\Avec+2m\left(\dot{\psf}^{(n:m)}||\nablav^{n-2m}_0
\right)\cdot\Delta^{m-1}_0(\nablav_0\cdot\Avec)\nonumber\\
\fl &+&\left.\left.(n-2m)\left(\msf^{(n:m)}_{\scr \rm sym}||
\nablav^{n-2m-1}\right)\cdot\Delta^m_0\Bvec\right]
-\frac{(n-1)}{n!}\left(\nablav^{n-2}||\Nsf^{(n-1)}\right)\cdot\left(\nablav_0
\times\Bvec\right)\right\}\nonumber
\eeqa

Because $\Delta\Avec=\nablav(\nablav\cdot\Avec)-\nablav\times\Bvec$, in the
 equations \eref{ww} and \eref{wmp} we may consider the relationships
\begin{equation}\label{rel1}
\fl\left[\dot{\Pi}^{(n-2)}||\nablav^{n-3}_0\right]\cdot\Delta_0\Avec=
\left[\dot{\Pi}^{(n-2)}||\nablav^{n-2}_0\right]\nablav_0\Avec-
\left[\dot{\Pi}^{(n-2)}||\nablav^{n-3}_0\right]\cdot\left(\nablav_0\times
\Bvec\right)
\end{equation}
and
\beqa\label{rel2}
\fl \left[\dot{\psf}^{(n:m)}||\nablav^{n-2m-1}\right]\cdot\Delta^m_0\Avec
&=&\left[\dot{\psf}^{(n:m)}||\nablav^{n-2m}\right]\Delta^{m-1}_0
\nablav_0\Avec\nonumber\\
&-&\left[\dot{\psf}^{(n:m)}||\nablav^{n-2m-1}\right]\cdot\Delta^{m-1}_0\left(\nablav\times\Bvec\right).
\eeqa
Using the equations \eref{rel1} and \eref{rel2} and considering also the equation 
$\Delta\Bvec=-\nablav\times(\nablav\times\Bvec)$ in the equations 
\eref{ww} and \eref{wmp} we may write
\beqa\label{www}
\fl W_{\scriptstyle\rm{int}}&=&{\mathcal W}_{\scriptstyle\rm{int}}
\nonumber\\
\fl&+&
\suml_{n\geq 2}\frac{n-1}{2n!}\left\{n
\left[\Pi^{(n-2)}||\nablav^{n-2}_0\right]\Delta_0\Phi
-n\left[\dot{\Pi}^{(n-2)}||\nablav^{n-2}_0\right]
\nablav_0\Avec\right.\nonumber\\
\fl &+&(n-2)\left[\dot{\Pi}^{(n-2)}||\nablav^{n-3}_0\right]
\cdot\left(\nablav_0\times\Bvec\right)-\frac{2}{n}\left[\nablav^{n-2}_0||
\Nsf^{(n-1)}\right]\cdot\left(\nablav_0\times\Bvec\right)\nonumber\\
\fl&+&\left.(n-2)\left[\bbox{\Lambda}^{(n-2)}||\nablav
^{n-3}_0\right]\cdot\nablav_0\times\left(\nablav_0\times\Bvec\right)\right\}
\eeqa
and
\beqa\label{wmppa}
\fl W_{\scr \rm int}&=&{\mathcal W}_{\scr\rm int}+\suml_{n\geq 2}
\frac{1}{n}\left\{\suml^{[n/2]}_{m=1}\frac{(-1)^m(2n-1-2m)!!}{(2n-1)!!
2^mm!(n-2m)!}\left[-n\left(\psf^{(n:m)}||\nablav^{n-2m}_0\right)
\Delta^m_0\Phi\right.\right.\nonumber\\
\fl &+&n\left(\dot{\psf}^{(n:m)}||
\nablav^{n-2m}_0\right)\Delta^{m-1}_0\nablav_0\Avec
-(n-2m)\left(\dot{\psf}^{(n:m)}||\nablav^{n-2m-1}_0
\right)\cdot\Delta^{m-1}_0(\nablav_0\times\Bvec)\nonumber\\
\fl&-&\left.(n-2m)\left(\msf^{(n:m)}_{\scr \rm sym}||
\nablav^{n-2m-1}\right)\cdot\Delta^{m-1}_0\nablav_0\times\left(\nablav_0
\times\Bvec\right)\right]\nonumber\\
\fl&-&\left.\frac{n-1}{n!}
\left[\nablav^{n-2}_0||\Nsf^{(n-1)}
\right]\cdot\left(\nablav_0\times\Bvec\right)\right\}
\eeqa
with $$\nablav_0\times\left(\nablav_0\times\Bvec\right)=\mu_0
\nablav_0\times\jvec_{\scr\rm ext}-\frac{1}{c^2}\frac{\d^2\Bvec}{\d t^2}.$$

\section{Concluding remarks}
Considering firstly the cases of the static external fields, we have  the 
separate electric and magnetic terms:
\begin{equation}\label{we}
W^{(\eps)}_{\scriptstyle\rm{int}}={\mathcal W}^{(\eps)}_{\scriptstyle\rm{int}}-
\frac{1}{\eps_0}\suml_{n\geq 2}\frac{n(n-1)}{n!}\left[\Pi^{(n-2)}||
\nablav^{n-2}_0\right]\rho_{\scriptstyle\rm{ext}}
\end{equation}
and
\beqa\label{wm}
W^{(\mu)}_{\scriptstyle\rm{int}}&=&{\mathcal W}^{(\mu)}_{\scriptstyle
\rm{int}}-\mu_0\suml_{n\geq 2}\frac{1}{n!}\left[\frac{n-1}{n}\left(
\Nsf^{(n-1)}||\nablav^{n-2}\right)\cdot\jvec_{\scriptstyle\rm{ext}}
\right.\nonumber\\
&-&\left.\frac{(n-1)(n-2)}{2}\left(\bbox{\Lambda}^{(n-2)}||\nablav^{n-3}_0\right)
\cdot\left(\nablav_0\times\jvec_{\scriptstyle\rm{ext}}\right)\right].
\eeqa
If the supports of the external sources do not intersect the supports 
of the given $(\rho,\jvec)$ distribution, then the interaction energies are 
invariant in respect to the substitutions of multipole tensors by 
the symmetric traceless ones. Differences appear when the intersection 
is not empty.
\par Denoting $W'_{\scr \rm{int}}=W_{\scr \rm{int}}-{\mathcal W}
_{\scr \rm{int}}$, it is easy to see that the gauge invariance of the 
 theory is satisfied separately by ${\mathcal W}_{\scr \rm{int}}$ 
and $W'_{\scr \rm{int}}$. \\
Let the external field potential satisfying the Lorenz constraint
\begin{equation}\label{Lorenz}
\nablav\cdot\Avec+\frac{1}{c^2}\frac{\d\Phi}{\d t}=0.
\end{equation}
Because in this case
\begin{equation}\label{KG}
\Delta\Avec=-\mu_0\jvec_{\scr \rm ext}+\frac{1}{c^2}\frac{\d^2\Avec}{\d t^2},\;\;
\Delta\Phi=-\frac{1}{\eps_0}\rho_{\scr \rm ext}+\frac{1}{c^2}\frac{\d^2
\Phi}{\d t^2},
\end{equation}
we may write
\beqa\label{WLorenz}
\fl W'_{\scr\rm int}&=&\suml_{n\geq 2}\frac{n-1}{2n!}\left\{-n\left[
\Pi^{n-2}||\nablav^{n-2}_0\right]\frac{1}{\eps_0}\rho_{\scr\rm ext}\right.
\nonumber\\
\fl  &+& (n-2)\left[\dot{\Pi}^{n-2}||\nablav^{n-3}\right]
\cdot\left(\nablav_0\times\Bvec\right)-(n-2)\left[\Lambda^{(n-2)}||
\nablav^{n-3}_0\right]\cdot\Delta_0\Bvec\nonumber\\
\fl &-&\left.\frac{2}{n}\left[\nablav^{n-2}||\Nsf^{(n-1)}\right]
\left(\nablav_0\times\Bvec\right)\right\}
-\frac{\d}{\d t}\left\{\suml_{n\geq 2}\frac{1}{2(n-2)!}\left[\Pi^{(
n-2)}||\nablav^{n-3}_0\right]\frac{\d \Phi}{\d t}\right\}.
\eeqa
We see that in the Lorenz gauge the potentials do not contribute 
actually to the part $W'_{\scr\rm int}$ of the interaction energy.
Moreover,  in the case of a free external field with the radiative gauge, 
$\nablav_0\cdot\Avec=0,\;\Phi=0$, the potential $\Avec$ contribute actually 
only in the part ${\mathcal W}_{\scr\rm int}$ of the interaction energy.\\
From the equation \eref{www}  we see also that in the Coulomb gauge for the
 external field the potentials do not cotribute to $W'_{\scr\rm int}$.

\vspace{0.5cm}
\par {\bf References}


\begin{thebibliography}{10}
\bibitem{cvsc}
Vrejoiu C. 1984 {\it St. Cercet Fiz.} {\bf 36} 863
\bibitem{gonzales}
Gonzales H, Juarez S R, Kielanowski P, Loewe M 1998 {\it Am.J.Phys.}, {\bf 
66} 228
\bibitem{vre}
Vrejoiu C  1993  {\it Electrodynamics and Relativity Theory}(in romanian)
 (E.D.P. Bucharest) 
\bibitem{cvjpa}
Vrejoiu C 2002 {\it J. Phys. A: Math. Gen.}, {\bf 35} 9911-22

\bibitem{cvdn}
Vrejoiu C, Nicmorus D 2003 {arXiv:physics/0307113 v1 23 Jul 2003}

\bibitem{jack}
Jackson J D 1975 {\it Classical Electrodynamics} (Wiley New York) 

\bibitem{castel}
Castellanos A, Panizo M, Rivas J 1978 {\it Am.J.Phys.}, {\bf 46} 1116-17
\bibitem{apple}
Applequist J. 1989 {\it J. Phys. A: Math. Gen.}, {\bf 22} 4303-4330 
\bibitem{jansen}
Jansen L 1957 {\it Physica} {\bf 23} 599

\end{thebibliography}
\end{document}